\input harvmac


\def\inbar{\,\vrule height1.5ex width.4pt depth0pt}
\font\cmss=cmss10 \font\cmsss=cmss10 at 7pt
\def\IZ{\relax\ifmmode\mathchoice
{\hbox{\cmss Z\kern-.4em Z}}{\hbox{\cmss Z\kern-.4em Z}}
{\lower.9pt\hbox{\cmsss Z\kern-.4em Z}} {\lower1.2pt\hbox{\cmsss
Z\kern-.4em Z}}\else{\cmss Z\kern-.4em Z}\fi}
\def\IB{\relax{\rm I\kern-.18em B}}
\def\IC{{\relax\hbox{$\inbar\kern-.3em{\rm C}$}}}
\def\ID{\relax{\rm I\kern-.18em D}}
\def\IE{\relax{\rm I\kern-.18em E}}
\def\IF{\relax{\rm I\kern-.18em F}}
\def\IG{\relax\hbox{$\inbar\kern-.3em{\rm G}$}}
\def\IGa{\relax\hbox{${\rm I}\kern-.18em\Gamma$}}
\def\IH{\relax{\rm I\kern-.18em H}}
\def\II{\relax{\rm I\kern-.18em I}}
\def\IK{\relax{\rm I\kern-.18em K}}
\def\IN{\relax{\rm I\kern-.18em N}}
\def\IP{\relax{\rm I\kern-.18em P}}
\def\IQ{\relax\hbox{$\inbar\kern-.3em{\rm Q}$}}
\def\IR{\relax{\rm I\kern-.18em R}}

\def\CA{{\cal A}}
\def\CF{{\cal F}}
\def\CJ{{\cal J}}
\def\CR{{\cal R}}
\def\CU{{\cal U}}
\def\CH{{\cal H}}

\def\CO{{\cal O}}
\def\CK{{\cal K}}

\def\a{\alpha}

\def\la{\lambda}

\def\th{\theta}
\def\bth{{\bar \theta}}

\def\bb{{\bar b}}

\def\bz{{\bar z}}
\def\pa{{\partial}}
\def\bpa{\bar{\partial}}
\def\ya{y_1}
\def\yb{y_2}

\def\za{z_1}
\def\zb{z_2}

\def\bza{\bar z_1}
\def\bzb{\bar z_2}
\def\bya{\bar y_1}

\def\ya{y_1}

\def\om{\omega}
\def\bom{\bar \omega}
\def\Om{\Omega}
\def\w{\wedge}
\def\tr{{\rm tr}}
\def\ln{{\rm ln}}

\def\w{\wedge}
\def\frac#1#2{{#1 \over #2}}

\def\bcs{{|c|^2}}
\def\bns{{|n|^2}}
\def\apas{{\frac{\alpha'}{a^2}}}

\lref\EH{
  T.~Eguchi and A.~J.~Hanson,
  ``Asymptotically flat selfdual solutions to Euclidean gravity,''
  Phys.\ Lett.\  B {\bf 74}, 249 (1978).
}

\lref\EHa{
  T.~Eguchi and A.~J.~Hanson,
  ``Selfdual solutions to Euclidean gravity,''
  Annals Phys.\  {\bf 120}, 82 (1979).
}

\lref\Bianchi{
  M.~Bianchi, F.~Fucito, G.~Rossi and M.~Martellini,
  ``Explicit construction of Yang-Mills instantons on ALE spaces,''
  Nucl.\ Phys.\  B {\bf 473}, 367 (1996)
  [arXiv:hep-th/9601162].
}

\lref\KN{ P.~B.~Kronheimer and H.~Nakajima,
 ``Yang-Mills instantons on ALE gravitational instantons,"
 Math.\ Ann.\ {\bf 288}, 263-307 (1990).
 }

\lref\CLP{
  M.~Cvetic, H.~Lu and C.~N.~Pope,
  Nucl.\ Phys.\  B {\bf 600}, 103 (2001)
  [arXiv:hep-th/0011023].
}

\lref\strom{ A.~Strominger, ``Superstrings with torsion,'' Nucl.\
Phys.\ B {\bf 274}, 253 (1986). }

\lref\DRS{K.~Dasgupta, G.~Rajesh and S.~Sethi, ``M theory,
orientifolds and G-flux,'' JHEP {\bf 9908}, 023 (1999)
  [arXiv:hep-th/9908088].}

\lref\BD{K.~Becker and K.~Dasgupta, ``Heterotic strings with
torsion,'' JHEP {\bf 0211}, 006 (2002)
  [arXiv:hep-th/0209077].}

\lref\GP{
  E.~Goldstein and S.~Prokushkin,
  ``Geometric model for complex non-K\"ahler manifolds with SU(3) structure,''
  Commun.\ Math.\ Phys.\  {\bf 251}, 65 (2004)
  [arXiv:hep-th/0212307].
}

\lref\FY{ J.-X.~Fu and S.-T.~Yau, ``The theory of superstring with
flux on non-K\"ahler manifolds and the complex Monge-Amp\`ere
equation,''  J.\ Differential Geom.\ {\bf 78}, 369  (2008) [arXiv:hep-th/0604063].
}

\lref\BBFTY{
  K.~Becker, M.~Becker, J.-X.~Fu, L.-S.~Tseng and S.-T.~Yau,
   ``Anomaly cancellation and smooth non-K\"ahler solutions in heterotic string theory,''
  Nucl.\ Phys.\ B {\bf 751}, 108 (2006)
  [arXiv:hep-th/0604137].}

\lref\stroma{
  A.~Strominger,
  ``Heterotic solitons,''
  Nucl.\ Phys.\  B {\bf 343}, 167 (1990)
  [Erratum-ibid.\  B {\bf 353}, 565 (1991)].
}

\lref\DX{
  M.~J.~Duff and J.~X.~Lu,
  ``Elementary five-brane solutions of D = 10 supergravity,''
  Nucl.\ Phys.\  B {\bf 354}, 141 (1991).
}

\lref\CHS{
  C.~G.~.~Callan, J.~A.~Harvey and A.~Strominger,
  ``World sheet approach to heterotic instantons and solitons,''
  Nucl.\ Phys.\  B {\bf 359}, 611 (1991).
}

\lref\CHSa{
  C.~G.~.~Callan, J.~A.~Harvey and A.~Strominger,
  ``Worldbrane actions for string solitons,''
  Nucl.\ Phys.\  B {\bf 367}, 60 (1991).
}

\lref\adams{
A.~Adams, Talk at {\it String Workshop on String Theory: From LHC Physics to Cosmology}, College Station, TX March 9-12, 2008, paper to appear.}

\lref\ael{
  A.~Adams, M.~Ernebjerg and J.~M.~Lapan,
  ``Linear models for flux vacua,''
 [arXiv:hep-th/0611084].
}

\lref\LVP{
  H.~Lu and J.~F.~Vazquez-Poritz,
 ``Resolution of overlapping branes,''
  Phys.\ Lett.\  B {\bf 534}, 155 (2002)
  [arXiv:hep-th/0202075].
}

\lref\GMW{
  J.~P.~Gauntlett, D.~Martelli and D.~Waldram,
  Phys.\ Rev.\  D {\bf 69}, 086002 (2004)
  [arXiv:hep-th/0302158].
}

\lref\CGLP{
  M.~Cvetic, G.~W.~Gibbons, H.~Lu and C.~N.~Pope,
  ``Ricci-flat metrics, harmonic forms and brane resolutions,''
  Commun.\ Math.\ Phys.\  {\bf 232}, 457 (2003)
  [arXiv:hep-th/0012011].
}

\lref\sen{
  A.~Sen,
  ``(2, 0) supersymmetry and space-time supersymmetry in the heterotic string
  theory,''
  Nucl.\ Phys.\  B {\bf 278}, 289 (1986).
}

\lref\ky{
  T.~Kimura and P.~Yi,
  ``Comments on heterotic flux compactifications,''
  JHEP {\bf 0607}, 030 (2006)
  [arXiv:hep-th/0605247].
}

\lref\BTY{
  M.~Becker, L.~S.~Tseng and S.~T.~Yau,
 ``Moduli space of torsional manifolds,''
  Nucl.\ Phys.\  B {\bf 786}, 119 (2007)
  [arXiv:hep-th/0612290].
}

\lref\hull{
  C.~M.~Hull,
  ``Anomalies, ambiguities and superstrings,''
  Phys.\ Lett.\  B {\bf 167}, 51 (1986);
}

\lref\fiuv{
  M.~Fernandez, S.~Ivanov, L.~Ugarte and R.~Villacampa,
  ``Non-K\"ahler heterotic string compactifications with non-zero fluxes and
  constant dilaton,''
  [arXiv:0804.1648].
}


\Title{ {\vbox{
}}} {\vbox{ \hbox{\centerline{Local Heterotic Torsional Models}}\hbox{} \hbox{\centerline{}} }}

\bigskip

\centerline{Ji-Xiang Fu$^{1}$, ~Li-Sheng
Tseng$^{2,3}$~and~Shing-Tung Yau$^2$}

\bigskip \bigskip

\centerline{$^1$ \it School of Mathematical Sciences, Fudan University, Shanghai 200433, P.R. China}

\medskip

\centerline{$^2$\it Department of Mathematics, Harvard University,
Cambridge, MA 02138, USA}

\medskip

\centerline{$^3$\it Center for the Fundamental Laws of Nature}

\centerline{\it Jefferson Physical Laboratory, Harvard University,
Cambridge, MA 02138, USA}

\bigskip
\bigskip

\bigskip

\centerline{\bf Abstract}

\bigskip

We present a class of smooth supersymmetric heterotic solutions with a non-compact Eguchi-Hanson space.  The non-compact geometry is embedded as the base of a six-dimensional non-K\"ahler manifold with a non-trivial torus fiber.  We solve the non-linear anomaly equation in this background exactly.   We also define a new charge that detects the non-K\"ahlerity of our solutions.

\bigskip
\baselineskip 18pt
\bigskip
\noindent

\Date{June, 2008}



\newsec{Introduction}

In this paper, we study six-dimensional supersymmetric
non-compact solutions of the ten-dimensional heterotic supergravity.   Non-compact
solutions can have different physical interpretations in string theory.  They may be
local models of a compact solution or they may correspond to the
supergravity descriptions of solitonic objects of the theory.

We demonstrate the existence of six-dimensional smooth solutions on $T^2$ bundles over an $ALE$ space. For the base being the minimally
resolved $\IC^2/\IZ_2$, we work out the solution in detail using the
Eguchi-Hanson metric \EH.  In solving this solution, we work in complex coordinates and exploit the $SU(2)$ global symmetry of the Eguchi-Hanson metric.  Importantly, the symmetry reduces the anomaly equation to a first-order non-linear differential equation which we solve exactly.

Our solutions are $1/2$ BPS and are asymptotically $\IR\IP^3 \times T^2$.  These local non-K\"ahler models are closely related to the compact heterotic models of $T^2$ bundle over $K3$ described in \refs{\FY,\BBFTY} (see also \refs{\DRS, \BD}).  
They give an explicit local description of the six-dimensional compact solution near an $A_1$ orbifold singularity of the base $K3$.  Moreover, it may be possible that our local solutions can  be consistently glued-in to resolve in a non-K\"ahler manner singular compact manifolds such as $T^4/\IZ_2 \times T^2$ or even $K3/\IZ_2 \times T^2$. 

Alternatively, the local solutions we construct can be interpreted to describe a heterotic five-brane that is wrapped around a torus and transverse to an Eguchi-Hanson space.   Heterotic five-brane solutions with a tranverse Eguchi-Hanson space \refs{\CLP, \CGLP} or wrapped over an $S^1$ \refs{\LVP, \GMW} have been discussed previously in the literature.  Solutions of this type differ from the original five-brane solution \refs{\stroma,\DX,\CHS,\CHSa} in that the five-brane charge can be sourced by a non-trivial $U(1)$ gauge field instead of an $SU(2)$ instanton.  Here, we point out that both the Eguchi-Hanson geometry and the non-trivial fibered torus induce non-trivial $H$ fluxes.  And of particular importance for the heterotic string is that their presence introduces highly non-linear terms in the anomaly differential equation.   A main purpose of this paper is to demonstrate that the induced fluxes can be carefully balanced to give smooth non-compact solutions that solve the heterotic supergravity exactly at one-loop order.

The outline of the paper is as follows.  In section 2, we review the supersymmetry conditions and the solution ansatz we will use.  In section 3, we write down explicitly the solution with an Eguchi-Hanson space and the differential equation that must be solved from the anomaly equation.  In section 4, we solve the differential equation exactly.  In section 5, we write down our solutions in general form and discuss their physical characteristics.  Though our smooth solutions have zero five-brane charge, they are in general non-zero under a new charge which we define that detects the non-K\"ahlerity of the solutions.

\newsec{Supersymmetry conditions and solution ansatz}

We start from the ten-dimensional heterotic supergravity on the product manifold, $M^{3,1} \times X^6$,  a four-dimensional Minkowski spacetime times a  six-dimensional manifold.  Preserving supersymmetry requires that $X^6$ is complex and has an $SU(3)$ holonomy with respect to a torsional connection.  The heterotic solution on $X^6$ can be described by a hermitian metric $J$, a holomorphic $(3,0)$-form $\Om$, and a stable gauge
bundle $E \subset SO(32) {\rm ~or~} E_8\times E_8$ with curvature
$F$.  The additional conditions from supersymmetry and the consistency of anomaly cancellation are
\eqn\balance{d(\|\Omega\|_J\, J\w J)=0~,}
\eqn\hermitym{F^{(2,0)}=F^{(0,2)}=0~,\quad F_{mn}J^{mn}=0~,}
\eqn\anomcan{2i\, \pa\bpa J = {\a' \over 4} [{\rm tr} (R\w R) - {\rm
tr}(F\w F)]~,}
where
\eqn\Omdef{i\,\Omega \w {\bar \Omega} = \frac{4}{3}\,\|\Omega\|_J^2 J\w J \w J~.}
 Following Strominger \strom, we take the curvature $R$ in \anomcan\
to be defined by the hermitian
connection.  Though the type of connection is not specified physically at one-loop order,\foot{Physical relationships between different connections have been discussed in \refs{\hull, \sen, \ky}.} the hermitian connection is the unique metric connection that is compatible with the complex structure and whose torsion tensor does not contain a (1,1) component.  Furthermore, the resulting ${\rm tr} (R\w R)$ is always a (2,2)-form.\foot{${\rm tr} (R\w R)$ for non-hermitian connections will generally contain (3,1) and (1,3) components.  Since the other two terms in the anomaly equation in \anomcan\ are (2,2)-forms, the presence of these additional components will likely over-constrain the system of differential equations as they must be set to zero.  We note that nilmanifold solutions with different connections have been discussed recently in \fiuv.}
  The above equations
define what is called the Strominger system in the mathematical literature.  It consists of a conformally balanced condition for the hermitian metric $J$, a
hermitian Yang-Mills condition for the bundle curvature $F$, and an
anomaly condition relating the difference of the two Pontryagin
classes, $p_1(R)$ and $p_1(F)$.  The relations to the physical fields -
the metric $g$, the antisymmetric three-form field $H$, and the
scalar dilaton field $\phi$ - are given as follows
\eqn\physrel{g_{mn}=J_{mr} I^r{}_n~,\qquad H=i(\bpa-\pa)J~,\qquad
e^{-2\phi}=\|\Omega\|_J~,}
where $I$ is the complex structure determined by the holomorphic three-form $\Omega$.

There is a much-studied solution ansatz on the $T^2$ bundle over a
Calabi-Yau two-fold \refs{\DRS, \BD, \GP, \FY, \BBFTY}.  The metric takes the form \eqn\metant{J= e^{u}
J_{CY_2} + \frac{i}{2} (dz + \beta)\w (d\bz + {\bar \beta})} where
$u$ is a function of the base Calabi-Yau and the torus curvature
$\om=d\th \equiv d(dz + \beta)$ satisfies the quantization and
primitivity conditions \eqn\torreq{\frac{\om}{2\pi\sqrt{\a'}} \in
H^{1,1}(M)\cap H^{2}(M,\IZ)~, \qquad\om \w J_{CY_2} = 0~.}  Taking
the holomorphic three-form to be $\Om^{3,0}= \Om^{2,0}_{CY_2} \w
\th$ which is a closed $(3,0)$-form by \torreq, it is straightforward to check that
the conformally balanced condition is satisfied for any function
$u$.   We note that with the metric and three-form ansatz, the conformal factor $e^u=e^{2\phi}$ which follows from the third equation of \physrel
\eqn\uphi{\|\Om\|^{-1}_J = e^u = e^{2\phi}~.}
Further, choosing a hermitian Yang-Mills curvature, $F$, pull-backed from the base $CY_2$, the anomaly equation \anomcan\ reduces to a non-linear second-order differential equation for $u$ (or equivalently the dilaton field) that must be solved.

Below, we analyze the case in which the base Calabi-Yau two-fold is taken
to be a non-compact $ALE$ space.  In particular, we shall work out
the case with the Eguchi-Hanson metric in detail.

\newsec{Eguchi-Hanson base solution}

Consider $\IC^2$ with coordinates $(\za, \zb)$ and an involution,
$\sigma: (\za,\zb) \to (-\za,-\zb)\,$. Let $M$ be the blow up of
$\IC^2/\sigma$ at the origin by a $\IP^1$. Then $M$ is biholomorphic
to $\CO_{\IP^1}(-2)=T^*\IP^1$, the cotangent bundle of $\IP^1$.  The
Eguchi-Hanson metric \refs{\EH, \EHa} is an explicit complete, smooth Ricci-flat metric
on $M$.

Outside the origin of $\IC^2/\sigma$, the metric is $SU(2)$
invariant and depends only on the radial coordinate $r^2 = |\za|^2 + |\zb|^2$.
Being K\"ahler, the metric can be expressed as
\eqn\Jeh{\eqalign{J_{EH} & = \frac{i}{2} \pa\bpa\, \CK(r^2) \cr &=
\frac{i}{2}\left[ k\, \pa\bpa r^2 + k' \pa r^2 \w \bpa r^2\right]} }
where the K\"ahler potential $\CK$, the function $k(r^2)=d\CK/dr^2$,
and its derivative $k'(r^2)=dk/dr^2$  are given by
\eqn\Kpot{\CK = \sqrt{r^4
+ a^4} + a^2 \log\left[ \frac{r^2}{\sqrt{r^4 + a^4} + a^2} \right]~,}
\eqn\kdef{ k = \sqrt{1+\frac{a^4}{r^{4}}}=
\frac{a^2}{r^2}\sqrt{1+\frac{r^4}{a^4}}~, \qquad k' =
-\frac{a^2}{r^4\sqrt{1+\frac{r^4}{a^4}}}~.} The constant $a > 0$ is
a measure of the diameter of the central $\IP^1$.

On $M$, there is a normalizable anti-self-dual closed $(1,1)$-form.
It corresponds to the curvature of the line bundle of the $\IP^1$
and has the form up to a constant $c$
\eqn\wdef{\eta= i  \pa\bpa\, \ln h = i  \left[ \frac{h'}{h}\, \pa\bpa r^2  +
\left(\frac{h'}{h}\right)' \pa r^2 \w \bpa r^2 \right]~.} The
function $(h'/h)$ can be found by imposing the primitivity
condition, $\om\w J_{EH} =0\,$.  This gives the differential equation
\eqn\hdifcon{ \frac{h'}{h} k + \left(\frac{h'}{h} k r^2\right)'
=0~,} which has the solution, modulo a multiplicative integration
constant,
\eqn\hdef{\frac{h'}{h}= \frac{1}{r^4 k}=\frac{1}{a^2
r^2\sqrt{1+\frac{r^4}{a^4}}}~,\qquad \left(\frac{h'}{h}\right)' =
-\,\frac{2\frac{r^4}{a^4} +1}{a^2r^4(1+\frac{r^4}{a^4})^{3/2}}~.}

We can now write down explicitly the $T^2$ bundle over the
Eguchi-Hanson space metric ansatz
\eqn\Jan{J= e^u J_{EH} + \frac{i}{2} \th \w \bth~.}
For the curvature of the torus bundle,
we utilized the anti-self-dual $(1,1)$-from,
\eqn\wdefa{\om = d\th = i c \,\pa\bpa\, \ln h = \frac{ic}{a^2} \left\{
\frac{1}{r^2\sqrt{1+\frac{r^4}{a^4}}} \pa\bpa r^2 -
\frac{2\frac{r^4}{a^4}  +1}{r^4(1+\frac{r^4}{a^4})^{3/2}} \pa r^2 \w
\bpa r^2\right\}~,}
having inserted \hdef\ into \wdef\ and allowed for an overall complex constant $c\,$.

The constant $c$ is quantized since $\frac{\om}{2\pi\sqrt{\a'}} \in
H^{1,1}(M)\cap H^{2}(M,\IZ)$. We can obtain the quantization
condition by integrating the curvature $\om$ over the $\IP^1$ at the
origin. Working in the coordinate chart ($\yb\neq 0$)
\eqn\coordsq{\ya= \frac{\za}{\zb}~, \qquad   \yb = \zb^2~,\qquad
r^2=|\za|^2+|\zb|^2= |\yb |(1+|\ya|^2)~,}  we integrate $\om$ over
$\IP^1$ parametrized by $\ya$ in the limit $\yb\to 0$.  We can
rewrite
\eqn\omexp{\om=\frac{ic}{a^2}\left\{\left[\frac{1}{(1+|\ya|^2)^2} +
\CO(|\yb|^2)\right] d\ya\w d\bya + \ldots \right\}~,}
where we have only written out only the $d\ya\w d\bya$ term.  Therefore,
\eqn\integp{\eqalign{\frac{1}{2\pi\sqrt{\a'}}\int_{\IP^1} \om  & =
\frac{1}{2\pi\sqrt{\a'}}\int
\,\frac{ic}{a^2}\frac{1}{(1+|\ya|^2)^2}\,d\ya\w d\bya \cr & =
\frac{1}{2\pi\sqrt{\a'}} \int_0^\infty \frac{2\pi c}{a^2}
\frac{dx^2}{(1+x^2)^2 }=  \frac{c}{a^2\sqrt{\a'}} ~.}}
The quantization requirement imposes
\eqn\cquan{c= a^2 \sqrt{\a'}\, n \equiv a^2 \sqrt{\a'} (n_1 + i n_2)~,\qquad  n_1,n_2 \in \IZ~.}



Having written down explicitly the metric which is conformally balanced by construction, we now proceed to discuss the gauge connection and the anomaly equation.

\subsec{Hermitian Yang-Mills connections and curvature}

By convention, our gauge curvature $F$ is imaginary and the
Hermitian Yang-Mills condition requires that it is also $(1,1)$
anti-self dual. $F$ takes value in the Lie algebra of $SO(32)$ or
$E_8\times E_8$.  Hermitian Yang-Mills connections on Eguchi-Hanson
space has been studied by Kronheimer and Nakajima for various rank
bundles.  In this paper, we will limit the discussion explicitly to the $U(1)$
case.


For the rank one or $U(1)$ gauge bundle, we note that there is only
the line bundle over $\IP^1$ so $F$ must be proportional to $\eta$
in \wdef.  In general, we can have a direct sum of $U(1)$ bundles.
The curvature for each $U(1)$ bundle takes the form \wdef
\eqn\curvdef{F= c' \pa\bpa\, \ln\, h = c' \left[ \frac{h'}{h}\, \pa\bpa
r^2  + \left(\frac{h'}{h}\right)' \pa r^2 \w \bpa r^2 \right]~, }
where $c'$ is a real number.  We then have
\eqn\Fsq{\eqalign{F\w F &
= c'^2 \left\{ \left(\frac{h'}{h}\right)^2 \pa\bpa r^2\w\pa\bpa r^2+
\left[\left(\frac{h'}{h}\right)^2\right]' \pa\bpa r^2 \w \pa r^2 \w
\bpa r^2\right\} \cr & \equiv \CF\, \pa\bpa r^2\w\pa\bpa r^2 + \CF'
\, \pa\bpa r^2 \w \pa r^2 \w \bpa r^2 }}
where
\eqn\fdef{\CF= c'^2\left(\frac{h'}{h}\right)^2 = \frac{c'^2}{a^4r^4(1+\frac{r^4}{a^4})}~.}
 The $U(1)$ gauge bundle also has a quantization:
$\frac{iF}{2\pi}\in H^{1,1}\cap H^{2}(\IZ)$.  Following the
computation in \omexp-\integp, this implies
\eqn\cpquan{c'= a^2\, m~, \qquad m \in \IZ~.}

\subsec{Anomaly equation}

With the metric ansatz \Jan, the anomaly equation is explicitly (see \FY\ for derivation)
\eqn\anom{\eqalign{2i\pa\bpa J &= \frac{\a'}{2}\left( \tr [R\w R] -
\tr [F\w F] \right ) \cr & = \frac{\a'}{2} \left( \tr [R_{EH}\w
R_{EH}]  + 2\, \pa\bpa u \w \pa \bpa u + 2\, \pa\bpa [e^{-u} \tr
(\bpa B\w \pa B^* \frac{g_{EH}^{-1}}{2}) ]  - \tr [F\w F] \right).}}
where $B$ is a column vector $B=\left(\matrix{B_1 \cr B_2}\right)$ locally defined such that $\bpa (B_1\, dz^1 + B_2\, dz^2 ) = \omega\,$.  Note that each term is a closed $(2,2)$-form on the base.  Since the
solution has $SU(2)$ global symmetry, we can express each term in terms
of a combination of $\pa\bpa r^2 \w \pa\bpa r^2$ and $\pa\bpa r^2 \w
\pa r^2 \w \bpa r^2$. We now proceed to calculate each term below.

\bigskip

\noindent{\it A. $dH=2i\pa\bpa J$ term}

Using \Jan\ for $J$, we find
\eqn\lhse{2i\pa\bpa J  =
2i\pa\bpa e^u\w J_{EH} - \om\w \bom ~,}
and
\eqn\fterm{2i\pa\bpa e^u \w J_{EH} = -\left[ (e^u)' k \,\pa\bpa
r^2\w\pa\bpa r^2 + [(e^u)' k ]' \pa\bpa r^2 \w \pa r^2 \w \bpa
r^2\right]~,}
\eqn\sterm{-\om \w \bom =  \bcs \left\{
\left(\frac{h'}{h}\right)^2 \pa\bpa r^2\w\pa\bpa r^2+
\left[\left(\frac{h'}{h}\right)^2\right]' \pa\bpa r^2 \w \pa r^2 \w
\bpa r^2\right\} ~.}
Combining the two terms, we can write
\eqn\lhsee{ 2i\pa\bpa J \, \equiv \CJ \pa\bpa
r^2\w\pa\bpa r^2 + \CJ' \,\pa\bpa r^2 \w \pa r^2 \w \bpa r^2~,}
where
\eqn\jdef{\CJ = - (e^u)' k + \bcs \left(\frac{h'}{h}\right)^2 = - (e^u)'\frac{a^2}{r^2}\sqrt{1+\frac{r^4}{a^4}} +  \frac{\bcs}{a^4r^4(1+\frac{r^4}{a^4})} ~.}
As will be needed shortly, we note here that $-\om\w \bom = \|\om\|^2 \frac{J^2_{EH}}{2!}$ implies
\eqn\onorm{\|\om\|^2 = -4 \bcs \left\{2
\left(\frac{h'}{h}\right)^2 +
\left[\left(\frac{h'}{h}\right)^2\right]' r^2 \right\}=
\frac{8\bcs}{a^8(1+\frac{r^4}{a^4})^2}~.}

\bigskip

\noindent{\it B.  $\tr [R_{EH}\w R_{EH}]$ term}

The curvature tensor is written in terms of metric
$(g_{EH})_{a\bb}=-i(J_{EH})_{a\bb}$ in $\Jeh$.  For the hermitian curvature, we find
\eqn\Rcal{\eqalign{R_{EH}&=\bpa((\pa g_{EH})\,g_{EH}^{-1})\cr &= \left[\frac{k'}{k} I - \frac{3\,k'}{r^2 k} M \right]  \bpa\pa r^2 +   \left[\left(\frac{k'}{k}\right)' I - \left(\frac{3\,k'}{r^2 k}\right)' M \right]\bpa r^2 \w \pa r^2 \cr  & \qquad+ \left(\frac{k'}{k}\right)' \bpa r^2 \w \pa M  - \left(\frac{3\,k'}{r^2 k}\right)' \bpa M \w \pa r^2 + \frac{k'}{k}\, \bpa\pa M~}}
with the $2 \times 2$ matrix $I=\delta_{ij}$  and  $M_{ij}= {\bar z}_i z_j\,$.  A long calculation results in
\eqn\trs{\eqalign{\tr [R_{EH}\w R_{EH}] &= 6 \left\{2
\left(\frac{k'}{k}\right)^2+
\left[\left(\frac{k'}{k}\right)^2\right]' r^2 \right\} d\za \w d\bza
\w d\zb \w d\bzb \cr & = 6 \left\{ \left(\frac{k'}{k}\right)^2
\pa\bpa r^2\w\pa\bpa r^2 + \left[\left(\frac{k'}{k}\right)^2\right]'
\pa\bpa r^2 \w \pa r^2 \w \bpa r^2 \right\}\cr  & \equiv \CR\,
\pa\bpa r^2\w\pa\bpa r^2 + \CR' \,\pa\bpa r^2 \w \pa r^2 \w \bpa r^2}}
with
\eqn\rsdef{\CR = 6 \left(\frac{k'}{k}\right)^2 =
\frac{6}{r^4 (1+\frac{r^4}{a^4})^2} ~.}
Alternatively, we can express
\eqn\trrexp{ \tr [R_{EH}\w R_{EH}]=
-\frac{24}{a^4(1+\frac{r^4}{a^4})^3} \,d\za\w d\bza \w d\zb \w
d\bzb~.}

\noindent {\it C. Other trace $R^2$ terms}

The $(\pa\bpa u)^2$ term can be formally written as
\eqn\usquar{\eqalign{2\pa\bpa u \w \pa\bpa u &= 2 \left\{(u')^2
\pa\bpa r^2 \w \pa \bpa r^2 + \left[(u')^2\right]' \pa\bpa r^2 \w
\pa r^2 \w \bpa r^2\right\} \cr & \equiv\CU\, \pa\bpa r^2\w\pa\bpa
r^2 + \CU' \,\pa\bpa r^2 \w \pa r^2 \w \bpa r^2}  }
where
\eqn\udef{\CU = 2(u')^2~.}

As for the remaining term, we use a formula in \FY
\eqn\trbsq{\eqalign{e^{-u}\tr (\bpa B\w \pa B^*
\frac{g_{EH}^{-1}}{2}) ] &= i \frac{\bcs}{4}\|\om\|^2 J_{EH}\cr &=-
e^{-u}\frac{\bcs}{(1+ r^4)^2}\left[ k\, \pa\bpa r^2 + k'\, \pa r^2
\w \bpa r^2 \right] \cr & \equiv \CH_1 \pa\bpa r^2  + \CH_2 \pa r^2
\w \bpa r^2} } where
\eqn\hhdef{\CH_1=
-e^{-u}\frac{\bcs}{a^6r^2(1+\frac{r^4}{a^4})^{3/2}}~,\qquad \CH_2 =
e^{-u}\frac{\bcs}{a^6r^4(1+\frac{r^4}{a^4})^{5/2}}~.}
This implies
\eqn\bbsq{\eqalign{2\, \pa\bpa [e^{-u} \tr (\bpa B\w \pa B^*
\frac{g_{EH}^{-1}}{2}) ]   &= 2\left\{ (\CH_1' - \CH_2) \pa\bpa r^2
\w \pa \bpa r^2  + (\CH_1' - \CH_2)' \pa\bpa r^2 \w \pa r^2 \w \bpa
r^2\right\} \cr &\equiv\CH\, \pa\bpa r^2\w\pa\bpa r^2 + \CH'
\,\pa\bpa r^2 \w \pa r^2 \w \bpa r^2}}
where
\eqn\hdef{\CH= 2(\CH_1'
- \CH_2) = 2 \bcs e^{-u}\left[
\frac{u'}{a^6r^2(1+\frac{r^4}{a^4})^{3/2}} +
\frac{4}{a^{10}(1+\frac{r^4}{a^4})^{5/2}}\right]~. }

\subsec{The resulting anomaly differential equation}

We can now write the anomaly equation \anom\ as
\eqn\anoms{\eqalign{2i\pa\bpa J - \frac{\a'}{2}\left( \tr [R\w R] -
\tr [F\w F] \right ) & \equiv \CA\, \pa\bpa r^2\w\pa\bpa r^2 + \CA'
\,\pa\bpa r^2 \w \pa r^2 \w \bpa r^2 \cr
& =\frac{1}{r^2}\left[\CA(r^2)\, r^4\right]'\, d\za\w d\bza \w d\zb \w d\bzb} }
where
\eqn\cadef{\CA= \CJ + \frac{\a'}{2}\CF - \frac{\a'}{2} ( \CR + \CU + \CH)~,}
written in terms of functions defined in
\jdef, \fdef, \rsdef, \udef, and \hdef.   The anomaly condition is therefore solved setting $\CA=0\,$.  With the quantization conditions \cquan\ and \cpquan, $\CA=0$ leads to the first order differential equation
\eqn\diffeqa{\eqalign{- u'\,
e^u\frac{a^2}{r^2}&\sqrt{1+\frac{r^4}{a^4}}  +
\frac{\a'\bns}{r^4(1+\frac{r^4}{a^4})} + \frac{\a' {m_i}^2}{2
r^4(1+\frac{r^4}{a^4})} \cr &=\a' \left[\frac{3}{r^4
(1+\frac{r^4}{a^4})^2} +  (u')^2 +  \a' \bns e^{-u}\left(
\frac{u'}{a^2 r^2(1+\frac{r^4}{a^4})^{3/2}} + \frac{4}{a^6
(1+\frac{r^4}{a^4})^{5/2}}\right) \right] } }
where \eqn\ndef{|n|^2 = n_1^2
+ n_2^2   \qquad {\rm and }\qquad   n_1,n_2, m_i \in \IZ~.}
In $m_i\,$, we have allowed for the possibility of multiple $U(1)$ gauge bundles
denoted by the index $i\,$.  Heterotic string allows for at most a rank 16 gauge bundle so $m_i^2$ should be taken to denote  $\sum_{j=1}^{16} m^2_j$.

For $\bns +\frac{{m_i}^2}{2}=3$, we
find that the differential equation has a smooth solution for $u$ for all values of $\apas>0$.  Explicitly, it takes the form
\eqn\solu{\eqalign{e^u&= \sum_{k=0}^\infty
\frac{a_k}{(1+\frac{r^4}{a^4})^{\frac{k}{2}}} \cr &= a_0\left[1 - \left(\frac{\alpha'}{a^2 a_0}\right)
\frac{1}{(1+\frac{r^4}{a^4})^{\frac{3}{2}}} + \left(\frac{\alpha'}{a^2 a_0}\right)^2
\frac{|n|^2}{(1+\frac{r^4}{a^4})^{{2}} }+ \left(\frac{\alpha'}{a^2 a_0}\right)^3
\frac{(|n|^2
+9/7)}{(1+\frac{r^4}{a^4})^{\frac{7}{2}}}+\ldots\right] }}
which converges for $\frac{1}{a_0}\apas < 1 $ sufficiently small.
In the next
section, we will derive the solution showing how the constants $a_k$
can be found iteratively and that the series converges to an exact
solution of the differential equation \diffeqa.

\newsec{Solving the anomaly equation}

To solve the differential equation, we first rewrite \diffeqa\ in a
more convenient form in a few steps.  To start, multiplying
\diffeqa\  by $1/a^{2}$ and re-arranging terms gives
\eqn\diffeqb{\eqalign{\frac{u'}{r^2}\,
e^u \sqrt{1+\frac{r^4}{a^4}}  + \left(\apas\right)&\left( u'^2  +
\frac{3 - (\bns +\frac{{m_i}^2}{2})}{r^4 (1+\frac{r^4}{a^4})^2} -
\frac{\bns +\frac{{m_i}^2}{2}}{a^4(1+\frac{r^4}{a^4})^2} \right)\cr  &
+ \left(\apas\right)^2 \bns e^{-u}\left(
\frac{u'}{r^2}\frac{1}{(1+\frac{r^4}{a^4})^{3/2}} +
\frac{4}{a^4(1+\frac{r^4}{a^4})^{5/2}}\right) = 0 }}
Setting
$\frac{\alpha'}{a^2}=\alpha$, $|n|^2+\frac{m_i^2}{2}=3$ and
replacing $u'\, e^u$ with $(e^u)'$, we find
\eqn\diffeqc{
\frac{(e^u)'}{r^2}\sqrt{1+\frac{r^4}{a^4}}-\alpha(e^u)'(e^{-u})'-\frac{3\alpha}{a^4(1+\frac{r^4}{a^4})^2}-\alpha^2|n|^2\frac 1
{r^2}\frac{(e^{-u})'}{(1+\frac{r^4}{a^4})^{\frac 3
2}}+4\alpha^2|n|^2\frac{e^{-u}}{a^4(1+\frac{r^4}{a^4})^{\frac 5
2}}=0~.}
And lastly, defining $e^u=v(s)\,,~ s=\frac{r^4}{a^4}\,$,
with $\frac{d}{dr^2} e^u = 2 \frac{\sqrt{s}}{a^2} \frac{d}{ds}v$ and multiplying through by $a^4 v^2$, we
arrive at the final form of the differential equation $D(\alpha, v)$ which we will solve
\eqn\definal{D(\alpha, v)= 2(1+s)^{\frac 1
2}v^2v'+4\alpha(1+s)v'^2-4\alpha
v'^2-\frac{3\alpha}{(1+s)^2}v^2+\frac{2\alpha^2|n|^2}{(1+s)^{\frac 3
2}}v'+\frac{4\alpha^2|n|^2}{(1+s)^{\frac 5 2}}v=0~.}

In writing $D(\alpha, v)$, we have emphasized the dependence of the differential equation on the parameter $\alpha$.  The solution function $v=v(s,\alpha)$ of course depends on the coordinate $s$ but should also vary with $\alpha$.  The presence of the parameter $\alpha$ is actually rather useful.  Together with $v$, we see that $D(\alpha, v)$ is indeed homogenous under the scaling
\eqn\scale{D(\la \alpha, \la v)= \lambda^3 D(\alpha, v)~,\qquad {\rm for~} \lambda \in
{\IR}^+~.}  This is important as it means that if we find a solution $D(\alpha_0, v_0)=0$ at a given value $\alpha=\alpha_0$, then for any other value $\alpha={\tilde \alpha}=\la \alpha_0$, there is also a solution given by $v= \la v_0$.  Taking advantage of this fact, we will solve $D(\alpha, v)$ for $\alpha<1$ and sufficiently small (which we shall make precise later).  The scaling of \scale\ then implies a solution for all $\alpha>0$.

The form of \definal\ suggests that we look for a
solution of the type
\eqn\vanstz{v=\sum_{k=0}^{\infty}\frac{a_k}{(1+s)^{\frac k 2}}~,}
with the coefficients $\a_k$'s possibly depending on the constants $\alpha$ and $|n|^2$.   Since the four-dimensional base metric in \metant\ should be asymptotic to the flat metric as $s\to\infty$, we must have $a_0> 0\,$.   This positive constant $a_0$ can be identified as a parameter of the solution space of $v(s,\alpha)$ for a given $\alpha$.\foot{From the string theory perspective, $a_0=e^{2 \phi_0}$ is the string coupling $g_s$ at the asymptotic infinity of the Eguchi-Hanson space.}   For notational simplicity, we shall set $a_0=1$ and find solutions for this case.   At the end of this section, we shall show how solutions with $a_0\neq 1$ can be easily obtained from those of $a_0=1$ via a scaling argument.

With the differential equation \definal\ and the solution ansatz $\vanstz$, we proceed  now to give a method to determine all the coefficients $a_k$.  We shall show that our prescription for the $a_k$'s results in $v$ being a convergent series for $\alpha$ sufficiently small.  We then prove that $v$ indeed converges to the solution $D(\alpha, v)=0$.

\subsec{Determining the coefficients $a_k$}

For specifying the $a_k$'s, we consider the finite series
\eqn\vpart{v_k= \sum_{l=0}^k \frac{a_l}{(1+s)^{\frac l 2}}~.}
We introduce the error function $E(v_k(s))=D(\alpha, v_k)$, or explicitly
\eqn\evk{ E(v_k)=2(1+s)^{\frac{1}{2}}v_k^2v_k'+4\alpha(1+s)v_k'^2-4\alpha
v_k'^2-\frac{3\alpha}{(1+s)^2}v_k^2+\frac{2\alpha^2|n|^2}{(1+s)^{\frac
3 2}}v_k'+\frac{4\alpha^2|n|^2}{(1+s)^{\frac 5 2}}v_k~.}
Thus for example,
\eqn\evo{
E(v_0)=-\frac{3\alpha}{(1+s)^2}+\frac{4\alpha|n|^2}{(1+s)^{\frac 5
2}}~. }
And making the choice $a_1=a_2=0$ and $a_3=-\alpha$ leads to
\eqn\evoa{E(v_0)=E(v_1)=E(v_2)~, }
and
\eqn\evoc{E(v_3)=\frac{4\alpha^2|n|^2}{(1+s)^{\frac 5
2}}-\frac{\alpha^3|n|^2-9\alpha^3}{(1+s)^4}-\frac{9\alpha^3}{(1+s)^5}~.}
Thus far, the error functions follow the form
\eqn\evke{E(v_k)=\frac{b_{k+2}}{(1+s)^{\frac{k+2}{2}}}+ \cdots~,}
with $b_{k+2}=0$ for $k=0,1,2\,$ and we have omitted terms of $\CO((1+s)^{-\frac{k+3}{2}})$.   In fact, we can iteratively choose $a_{k+1}$ such that \evke\ also holds for any $k>3$.   To show this, we first write
\eqn\evko{E(v_{k+1})=E(v_k)+\bigl(E(v_{k+1})-E(v_k)\bigr)~. }
We observe that
\eqn\evkd{
E(v_{k+1})-E(v_k)=\frac{-(k+1)a_{k+1}}{(1+s)^{\frac{k+2}{2}}}+\cdots~,}
which comes from the first term $2(1+s)^{\frac 12}v_k^2v'_{k+1}$ in \evk.  Comparing \evke\  and \evkd, we can set
\eqn\akde{ a_{k+1}=\frac{b_{k+2}}{k+1}~, }
which would cancel the $\frac{b_{k+2}}{(1+s)^{\frac{k+2}{2}}}$ term and gives us for \evko
\eqn\label{E(v_{k+1})=\frac{b_{k+3}}{(1+s)^{\frac{k+3}{2}}}+\cdots~. }
We shall choose each $a_k$'s similarly and thereby ensure \evke\ is valid for all $k$.

We have thus given an algorithm to determine each $a_k$ from those $a_i$'s with $i<k$.  Explicitly, the coefficients are given by
\eqn\akdef{\eqalign{
  a_{k+1}=&\frac 1 {k+1}
\Bigl\{-\alpha^2|n|^2(k-7)a_{k-3}-3\,\alpha\sum_{i,j=0}^{k-2}\sum_{i+j=k-2}a_ia_j
-\alpha\sum_{i,j=1}^{k-3}\sum_{i+j=k-2}ij\,a_ia_j\cr &\ \ \ \ \ \ \ \
+\alpha\sum_{i,j=1}^{k-1}\sum_{i+j=k}ij\,a_ia_j-\sum_{i,j=0}^{k}\sum_{l=1}^{k}\sum_{i+j+l=k+1}l\,a_ia_ja_l\Bigr\}
} } Using this formula, we find for instance
\eqn\aval{\eqalign{
a_4&=\alpha^2|n|^2\,,\ a_5=0\,, \ a_6=0\,,\
a_7=\alpha^3\left(|n|^2+ \frac{9}{7} \right),\cr
a_8&=-\alpha^4 \left(|n|^4+ 3|n|^2\right),\ a_9= \alpha^3\left(-1 +  \frac{16}{9}\alpha^2 |n|^4\right) \,,} }
and so on.

\subsec{Estimates for $a_k$ and convergence}

Being able to iteratively generate the coefficients of each term of
the series \vanstz, we can now show that the series converges when
$\alpha<1$ is sufficiently small.   Since $|a_3|=\alpha<1$ is small,
we can write
\eqn\atest{ |a_3|=\frac {\alpha_0^3}{3^3\,C}~, }
for some large constant $C$ and small $\alpha_0<1$. For a fixed
$\alpha_0<1$ and with \akdef\ and \atest, we shall prove by induction that
when $C$ is sufficiently large,
\eqn\akest{ |a_k|\leq \frac{\alpha_0^k}{k^3\, C }~.}
This estimate then immediately implies that the series $\sum_{k=0}^\infty
\frac{a_k}{(1+s)^{\frac k 2}}$ converges for any $s\geq 0$ since
$\alpha_0 < 1\,$.  We proceed now with the induction proof of \akest.

Let us assume that \akest\ is true for $1\leq k\leq N$ and
$N\geq 3$.  We shall prove that \akest\ is then also true for
$k=N+1$.  We show this by deriving explicit estimates for all five terms in the expression for $a_k$ in \akdef\ for $k=N+1$.  As convention, we take as definition $0^k=1$ below.

Starting with the first term of \akdef, we find the estimate
\eqn\first{\eqalign{\frac{\mid\alpha^2|n|^2(N-7)a_{N-3}\mid}{N+1}&\leq
\frac{\alpha^2|n|^2|N-7|}{N+1}\frac{\alpha_0^{N-3}}{(N-3)^3C}\cr
&\leq
\frac{\alpha_0^{N+1}}{(N+1)^3C}\frac{\alpha_0^2}{C^2}\frac{|n|^2|N-7|(N+1)^2}{3^6(N-3)^3}\cr
&\leq \frac{\alpha_0^{N+1}}{(N+1)^3C}\frac{\alpha_0^2}{C^2} C_1}}
with the constant
\eqn\c{C_1= \sup_{i\geq 3}\frac{|n|^2|i-7|(i+1)^2}{3^6(i-3)^3}~.}
For the estimate of the second term in \akest\ for $k=N+1$, we find
\eqn\second{\eqalign{\frac{3\alpha}{N+1}\sum_{i+j=N-2}\sum_{i,
j\geq 0}\mid a_ia_j\mid\ \leq&
\frac{3\alpha}{N+1}\frac{\alpha_0^{N-2}}{C^2}\sum_{i+j=N-2}\sum_{i,j\geq
0}\frac 1{i^3j^3}\cr \leq&
\frac{\alpha_0^{N+1}}{(N+1)C}\frac{6\alpha\alpha_0^{-3}}{C}\sum_{j=N-2-i}\frac
1 {j^3}\sum_{i\geq [\frac {N-2}{2}]}^{N-2}\frac{1}{i^3}\cr \leq
&\frac{\alpha_0^{N+1}}{(N+1)^3C}\frac{1}{C^2}\frac{16(N+1)^2}{9(N-2)^3}\sum_{j=0}^{N-2-[\frac{N-2}{2}]}\frac
1 {j^3}\cr \leq &\frac{\alpha_0^{N+1}}{(N+1)^3C}\frac{C_2}{C^2}}}
with the constant \eqn\cc{C_2=\frac{2^8}{9}\sum_{j=0}^{\infty}\frac 1
{j^3}~.}
The estimates for the third and fourth term are found similarly.   For the third term, we find
\eqn\third{\frac{\alpha}{N+1}\sum_{i+j=N-2}\sum_{i,j\geq 1}ij\mid
a_ia_j\mid\leq \frac{\alpha_0^{N+1}}{(N+1)^3C}\frac{C_3}{C^2}}
with the constant
\eqn\ccc {C_3=\frac{2^6}{3^3}\sum_{j=1}^{\infty}\frac 1 {j^2}~,}
and for the fourth term
\eqn\forth{\frac{\alpha}{N+1}\sum_{i+j=N}\sum_{i,j\geq 1}ij\mid
a_ia_j\mid\leq \frac{\alpha_0^{N+1}}{(N+1)^3C}\frac{C_4}{C^2}}
with the constant
\eqn\cccc{C_4=\frac{2^6}{3^5}\sum_{j=1}^{\infty}\frac{1}{j^2}~.}
Lastly, we estimate the fifth term in \akdef\ for $k=N+1$.  From
direct calculation, we obtain
\eqn\fifth{\eqalign{&\frac 1 {N+1}\sum_{i,j\geq 0,l\geq 1}\sum_{i+j+l=N+1}l\mid a_ia_ja_l\mid \cr \leq &\frac{\alpha_0^{N+1}}{(N+1)C^3}\sum_{i,j\geq 0,l\geq
1}\sum_{i+j+l=N+1}\frac{1}{l^2i^3j^3}\cr \leq
&\frac{\alpha_0^{N+1}}{(N+1)C^3}\sum_{i+j=N-1-l}\Bigl(\sum_{l\geq[\frac{N+1}{3}]}+\sum_{l<[\frac{N+1}{3}]}\Bigr)
\frac{1}{l^2i^3j^3}\cr \leq&
\frac{9\alpha_0^{N+1}}{(N+1)^3C^3}\Bigl(\sum_{i+j=0}^{N+1-[\frac{N+1}{3}]}\frac{1}{i^3j^3}+2\sum_{j=N+1-i-l}\sum_{i\geq
[\frac{N+1}{3}]}\sum_{l\leq[\frac{N+1}{3}]}\frac 1
{l^2ij^3}\Bigr)\cr \leq
&\frac{27\alpha_0^{N+1}}{(N+1)^3C^3}\sum_{i+j=0}^{N+1-[\frac{N+1}{3}]}\frac
1 {i^3j^3}\cr \leq &\frac
{\alpha_0^{N+1}}{(N+1)^3C}\frac{C_5}{C^2}}}
with the constant
\eqn\ccccc{C_5=27\sum_{i+j=0}^{\infty}\frac 1 {i^3j^3}~.}

Now let $C_0=\max\{C_1,C_2,C_3,C_4,C_5\}$. For $\alpha_0<1\,$, we choose the constant
\eqn\cchoice{C\geq\sqrt{5\,C_0}~.}
By summing over the five estimates in \first, \second,
 \third, \forth, and \fifth, we obtain the estimate
 \eqn\aNest{|a_{N+1}|
 \leq \frac{\alpha_0^{N+1}}{(N+1)^3C}~.}
And by induction, we have proven the desired estimate \akest\ and therefore
$v(s)=\sum_{k=0}^{\infty}\frac{a_k}{(1+s)^{\frac k 2}}$ converges for any $s$.

Having shown that the series $v$ converges, we still need to make
sure that $v=e^u>0$.  This positivity condition will give us a bound
on $\alpha$ for solutions with $a_0=1$.  Clearly for any $s\geq 0$,
\eqn\vpos{v> 1-\frac 1 {3^3C}\sum_{k\geq 3}\alpha_0^k =
1-\frac{\alpha_0^3}{3^3(1-\alpha_0)C}=1-\frac{\alpha}{1-\alpha_0}~.}
Since, $0<\alpha_0<1$, \vpos\ gives the condition
\eqn\alphaz{\alpha\leq 1-\alpha_0<1} to ensure $v(s)>0\,$.  Let
$\tilde\alpha>0$ be the solution of the equation
\eqn\alphaine{\tilde\alpha=\frac{(1-\tilde\alpha)^3}{3^3\sqrt{5\,C_0}}~.}
Then by \atest, \cchoice, and \alphaz, for any
$0<\alpha\leq\tilde\alpha$,
 $v(s)=\sum_{k=0}^{\infty}\frac{a_k}{(1+s)^{\frac k 2}}$
converges and $v(s)>0$ for all $s\geq 0\,$.

\subsec{Proving the series solves the differential equation}

Finally, having established that $v$ is a convergent series, we now
prove that $v$ is indeed a solution to the differential equation
\definal.  This is equivalent to showing that the error vanishes for
the entire series, {\it i.e.}
\eqn\elimit{\lim_{k\rightarrow\infty}E(v_k)=0~.} Since the leading
term is $(1+s)^{-\frac{k+2}{2}}$, we can write
\eqn\label{E(v_k)=\sum_{p=k+2}^{\frac{3k+1}{2}}\frac{c_p}{(1+s)^{\frac
p 2}}~, } with $c_{k+2}=b_{k+2}$.  By direct computation, we find
\eqn\cpdef{\eqalign{
c_{p}=&-\alpha^2|n|^2(p-9)a_{p-5}-3\alpha\sum_{i,j=0}^{k}\sum_{i+j=p-4}a_ia_j
-\alpha\sum_{i,j=1}^{k}\sum_{i+j=p-4}ij\,a_ia_j\cr &\ \ \ \ \ \ \ \
+\alpha\sum_{i,j=1}^{k}\sum_{i+j=p-2}ij\,a_ia_j-\sum_{i,j=0}^{k}\sum_{l=1}^{k}\sum_{i+j+l=p-1}l\,a_ia_ja_l
}}
 and the first term is zero if $p>k+5$. Similar to the
estimate for $|a_k|$ in \akest, we find the estimate for $|c_p|$
\eqn\label{ |c_p|\leq C(p-1)\parallel a_{p-1}\parallel\leq
\frac{\alpha_0^{p-1}}{(p-1)^2}~, } where we denote $\parallel
a_{p-1}\parallel$ the summation  of absolute values of every term in
$a_{p-1}$. Therefore, \eqn\label{ |E(v_k)|\leq
\sum_{p=k+2}^{\frac{3k+1}{2}}\frac{|c_p|}{(1+s)^{\frac p
2}}\leq\frac{\alpha_0^{k+1}}{(1+s)^{\frac{k+2}{2}}}\sum_{p=k+2}^{\frac{3k+1}{2}}\frac{1}{(p-1)^2}\rightarrow
0~, } as $k\to \infty$.  This proves $E(v)=0\,$.

\subsec{Solution and parameter space}

We have shown that the differential equation $D(\alpha,v)=0$ in \definal\ is solved by the convergent series
\eqn\vexp{v(s, \alpha)= \sum_{k=0}^\infty \frac{a_k}{(1+s)^{\frac{k}{2}}} = 1 - \frac{\alpha}{(1+s)^{\frac{3}{2}}} + \frac{\alpha^2\, |n|^2}{(1+s)^2} + \frac{\alpha^3\left(|n|^2+ \frac{9}{7} \right)}{(1+s)^{\frac{3}{2}}} + \cdots ~, }
for $\alpha\leq\tilde\alpha$ and $a_k$ given by \akdef.

We can now use the scale invariance of  $D(\alpha,v)=0$ in \scale\ to demonstrate a one parameter family of solution for any given value of $\alpha$.  We first show this for $\alpha=\tilde\alpha\,$ as defined in \alphaine\ for $a_0=1$ solutions.   Let $\alpha_0<\tilde\alpha$ and write $\alpha_0= \tilde\alpha/\la$ for a real constant $\la>1$.  At $\alpha=\alpha_0$, we have the solution $v(s,\alpha_0)$ given in \vexp.  Making use of the scaling of \scale, we obtain
\eqn\scales{0=D(\alpha_0, v(s, \alpha_0))=D(\frac{1}{\la} \tilde\alpha, \frac{1}{\la} \la \,v(s,\frac{\tilde\alpha}{\la}))= \frac{1}{\la^3} D(\tilde\alpha, \la\, v(s, \frac{\tilde\alpha}{\la}))~.}
This implies a family of solutions parametrized by $\la$ at $\alpha=\tilde\alpha$ given by
\eqn\vscal{v_\la(s,\tilde\alpha)= \la \,v(s, \frac{\tilde\alpha}{\la}) = \la\left[1 - \frac{\tilde\alpha}{\la}\frac{1}{(1+s)^{\frac{3}{2}}} + \left(\frac{\tilde\alpha}{\la}\right)^2 \frac{|n|^2}{(1+s)^2} + \left(\frac{\tilde\alpha}{\la}\right)^3\frac{\left(|n|^2+ \frac{9}{7} \right)}{(1+s)^{\frac{3}{2}}} + \cdots \right] ~,}
with $\la=[1,\infty)\,.$   To show a family of solutions for any value of $\alpha=\mu\,\tilde\alpha$ for any real constant $\mu$, we apply the scaling of \scale\ again to obtain
\eqn\vgen{v_\la(s,\alpha)= \mu\, v_\la(s,\tilde\alpha) = \mu\la\, v(s,\frac{\alpha}{\mu\la})}
In terms of the original expansion $v=\sum_{k=0}^\infty \frac{a_k}{(1+s)^{\frac{k}{2}}}\,$, we find that $a_0=\mu\la$ and we have convergence to a solution for $a_0=[\mu, \infty)\,$.  More simply, we write the convergent solution as
\eqn\vfinal{v(s,\alpha)=a_0 \left[1 - \frac{\alpha}{a_0}\frac{1}{(1+s)^{\frac{3}{2}}} + \left(\frac{\alpha}{a_0}\right)^2 \frac{|n|^2}{(1+s)^2} + \left(\frac{\alpha}{a_0}\right)^3\frac{\left(|n|^2+ \frac{9}{7} \right)}{(1+s)^{\frac{3}{2}}} + \cdots \right] ~,}
with the condition $\frac{\alpha}{a_0} = \frac{\tilde\alpha}{\la} < 1$ sufficiently small (since $\tilde\alpha <1$ and $\la\geq 1$).

In summary, we have found a one-parameter family of solutions for the anomaly equation for any value of $\alpha=\alpha'/a^2$

\newsec{Discussion}

We have constructed a class of smooth non-compact solutions that exactly solve the heterotic supergravity supersymmetry constraints to first order in $\alpha'$.  We write below the solution in the most general form, introducing the complex moduli $\tau= \tau_1 + i \tau_2$ (in $z=x+ \tau y$) and area $A$ of the torus as parameters:
\eqn\jsum{J=e^{u} \, J_{EH} + \frac{i}{2} \frac{A}{\tau_2} (dz + \beta ) \wedge (d\bz + {\bar \beta})~,}
\eqn\jehsum{J_{EH} = \frac{i}{2} \left[ \frac{a^2}{r^2}\sqrt{1 + \frac{r^4}{a^4}} \pa\bpa r^2  - \frac{a^2}{r^4} \frac{1}{\sqrt{1 + \frac{r^4}{a^4}}} \pa r^2 \wedge \bpa r^2 \right] }
\eqn\omsum{\om=d\beta =  i \sqrt{\alpha'} (n_1 + \tau n_2) \left[ \frac{1}{r^2 \sqrt{1 + \frac{r^4}{a^4}}} \pa \bpa r^2 - \frac{2 \frac{r^4}{a^4} + 1 } {r^4 (1 + \frac{r^4}{a^4})^{\frac{3}{2}}} \pa r^2 \w \bpa r^2 \right]}
\eqn\Fsum{F_i = m_i  \left[\frac{1}{r^2 \sqrt{1 + \frac{r^4}{a^4}}} \pa \bpa r^2 - \frac{2 \frac{r^4}{a^4} + 1 } {r^4(1 + \frac{r^4}{a^4})^{\frac{3}{2}}} \pa r^2 \w \bpa r^2 \right]}
\eqn\usum{\eqalign{e^{2\phi}&=e^u= \sum_{k=0}^\infty \frac{a_k}{(1+\frac{r^4}{a^4})^{\frac{k}{2}}}\cr  &=  e^{2\phi_0}\left[ 1 - \frac{\alpha'}{e^{2\phi_0}a^2}\frac{1}{(1+ \frac{r^4}{a^4})^{\frac{3}{2}}} +  \left(\frac{\alpha'}{e^{2\phi_0}a^2}\right)^2 \frac{\frac{A}{\tau_2} |n_1 + \tau n_2|^2}{(1+ \frac{r^4}{a^4})^2} + \cdots \right] }}
for
\eqn\constsum{ \frac{A}{\tau_2} |n_1 + \tau n_2|^2 + \frac{{m_i}^2}{2} =3 ~, \qquad {\rm and }\qquad   n_1,n_2, m_i \in \IZ~.}
and $e^{\phi_0}$ is the string coupling at asymptotic spatial infinity $r\to \infty$.   From \omsum\ and \Fsum, we see that both the torus twist curvature $\omega$ and the $U(1)$ gauge fields curvature $F$ are localized around the origin of the Eguchi-Hanson space and vanish in the asymptotic limit of $r\to \infty$.   The expression for $e^{2\phi}$ in \usum\ is obtained from \vfinal\ by replacing $|n|^2 \to \frac{A}{\tau_2} |n_1 + \tau n_2|^2$ and setting $a_0= e^{2\phi_0}$.  The condition for the convergence of the series then becomes
\eqn\agcom{\left(\frac{\alpha'}{a^2}\right)\frac{1}{g_s^2} < 1~,}
and sufficiently small.   
Clearly, our solution is consistent in the supergravity limit of $g_s \ll 1$ and $\alpha/a^2 \ll 1\,$ for sufficiently large $a^2$.

We observe that our solution with non-zero $H$ fluxes have moduli which may be constrained but are not fixed. Certainly the string coupling, $g_s=e^{\phi_0}\,$, and the size of the resolved $\IP^1$ as measured by $a^2$
 are not fixed.  Together, they are constrained by \agcom.  As for the torus, equation \constsum\ gives only
 one constraint for the torus area $A$ and complex structure moduli $\tau$ combined.  Thus, we are free to
 vary $\tau$ with a compensating variation of $A$.\foot{In the compact case of $T^2$ bundle over $K3$ base as discussed in \BTY, the torus complex structure moduli can be fixed with appropriately chosen $\omega=\omega_1 + \tau \omega_2 \in H^{2,0} \oplus H^{1,1}\,$. Here, the Eguchi-Hanson base is special in that it has only one normalizable two-form.}  Nevertheless, if $n_1$ and $n_2$  are not both zero, the area of the torus is constrained to be of $\CO(\alpha')$ (as $A$ is normalized with respect to $\alpha'$ in \constsum),



If we treat our solution as a solitonic object, we should determine its five-brane charge.  This charge can be obtained by integrating $H=d^c J$ at the spatial infinity of the transverse Eguchi-Hanson space, $EH$.   However, because of the non-trivial fibering, the Eguchi-Hanson space is not a four-dimensional submanifold of $X^6\,$ and so taking the spatial infinity limit of $EH$ is ill-defined in $X^6$.  Thus, to be rigorous,  we should pull-back $\IR\IP^3(r)$ at the radial coordinate $r$ in $EH$ to a $T^2$ bundle over $\IR\IP^3(r)$ which is a submanifold over $X^6$.  Denoting this five-submanifold by $S(r)$, we define the five-brane charge in $X^6$ as\foot{For simplicity, we have set $A=1$ and $\tau=i$ for the moduli of the torus in the discussion.  The area of the torus is conventionally normalized to $(2\pi \sqrt{\alpha'})^2\,$.}   
\eqn\fivex{\eqalign{Q_5 &= \lim_{r\to \infty}\frac{1}{(4\pi^2 \alpha')^2   } \int_{S(r)}H\wedge J =  \lim_{r\to \infty}\frac{1}{(4\pi^2 \alpha')^2 } \int_{S(r)} H \wedge \left(\frac{i}{2} \th\w {\bar \th}\right)\cr &= \lim_{r\to \infty}\frac{1}{4\pi^2 \alpha'} \int_{\IR\IP^3(r)} H =  \lim_{r\to \infty} \frac{1}{4\pi^2 \alpha'} \int _{\IR\IP^3 (r)} i(\bpa -\pa) e^u \wedge J_{EH}\cr
& = -\lim_{r\to \infty}\frac{1}{8\pi^2 \alpha'}\int _{\IR\IP^3(r)}(e^u)' \frac{a^2}{r^2}\sqrt{1+ \frac{r^4}{a^4}} (\bpa r^2 \w \pa\bpa r^2 + \pa r^2 \w \bpa\pa r^2)\cr
&=-\lim_{r\to \infty}\frac{1}{2\alpha'}\left[ r^4 (e^u)'  \sqrt{1+\frac{a^4}{r^4}} \,\right] }}
having used \jsum\ and \jehsum.   Plugging in the expression for $e^u$ in \usum, we find that the total net charge is zero.  This is perhaps as expected since in imposing the condition \constsum, we have effectively cancelled the negative charge contribution from the curvature of the Eguchi-Hanson space with the positive charge contribution from the torus twist and gauge fields.  A non-zero five-brane charge would likely require a singular solution.

Being zero, the five-brane charge can not distinguish between different torus curvature $\omega$  which when non-zero makes $X^6$ a non-K\"ahler manifold.  We can however define a new charge 
\eqn\qtdef{\eqalign{{\tilde Q} &= \frac{1}{(4\pi^2 \alpha')^2 } \int_{X^6}  dH \w J \cr  & = \frac{1}{(4\pi^2 \alpha')^2 } \int_{X^6} \left(2i\pa\bpa e^u \w J_{EH}-\omega \w {\bar \omega}\right) \w \left(\frac{i}{2} \theta \w {\bar \theta}\right)}}
where we have used the primitivity condition $\omega \w J_{EH} =0\,$.  Now, the first term on the right hand side, integrates to zero since it is a total derivative with zero boundary contribution as in \fivex.  The second term reduces to an integral on $EH$
\eqn\qtdeff{{\tilde Q}=-\frac{1}{4\pi^2 \alpha'}\int_{EH} \omega \w {\bar \omega}= -\frac{1}{4\pi^2\alpha'}\int_{\IC^2\!\!/\IZ_2} \frac{2 \,\alpha'|n|^2}{a^4(1+\frac{r^4}{a^4})^2} dz_1 \w d\bza \w d\zb \w d\bzb = \frac{1}{2}|n|^2~.}
Therefore, when $\om\neq 0$ and $X^6$ non-K\"ahler, $\tilde Q \neq 0$.  

A motivation for considering the charge $\tilde Q$ is that for the K\"ahler case where $dJ=0$, Stokes's theorem implies $\tilde Q=Q_5$ (compare \fivex\ with \qtdef).  Note that $dH$ corresponds to the source density of the five-brane.  But when $J$ is not K\"ahler, we have
\eqn\qq{{\tilde Q} - Q_5 = \int_{X^6} H \w dJ = -2i \int_{X^6} \pa J \w \bpa J~,}    
Hence, the difference between $\tilde Q$ and $Q_5$ implies non-K\"ahlerity.  
We also note that for the compact case, $\tilde Q$ is well-defined for $J$ as a class in the $\pa\bpa$ cohomology.  That is, $\tilde Q$ is invariant under $J\to J + \pa {\bar \gamma} + \bpa {\gamma}$ where $\gamma$ is $(1,0)$-form.  This may be relevant as the anomaly equation \anomcan\ is locally a $\pa\bpa$ equation \BTY.


It is expected that as higher order $\alpha'$ corrections to the supergravity constraints are taken into account,
the explicit form of our solutions will be corrected.  The explicit form as in the series expansion of \solu\ suggests that the corrections can probably be incorporated order by order in $\alpha'$.  Alternatively, one would like to have a worldsheet conformal field theory description of the geometrical model.  Such has been presented in \adams\ using the gauged linear sigma model formalism of \ael.

We have given a detailed study of the solution of a torus bundle over a non-compact Eguchi-Hanson space with $U(1)$ gauge bundles.  This can be considered the simplest case of a more general class of solutions that involve non-Abelian gauge bundles and more general $ALE$ base geometry.  Investigations on these more general solutions are interesting and we plan to report on them elsewhere.

\bigskip\bigskip\bigskip\bigskip
\centerline{\bf Acknowledgements}
\medskip
We thank A.~Adams, M.~Becker, J.~Harvey, J.T.~Liu, A.~Strominger, and A.~Tomasiello for discussion.  J.-X.~Fu is supported in part by NSFC grant 10771037, LMNS, and Fan Fund.  
L.-S.~Tseng is supported in part by NSF grant PHY-0714648. And S.-T.~Yau
is supported in part by NSF grants DMS-0306600 and PHY-0714648.

\listrefs

\end